\documentstyle[twoside,fleqn,espcrc2,emlines,bezier]{article}

\newcommand{\tr}[1]{\,{\rm tr}\,#1\,}
\newcommand{\NP}[1]{Nucl. \ Phys.}
\newcommand{\PL}[1]{Phys. \ Lett.}
\newcommand{\p}[1]{\partial}
\newcommand{\PRL}[1]{Phys.\ Rev.\ Lett. }
\newcommand{\MPL}[1] { Mod. Phys. Lett. }
\newcommand{\IJMP}[1] { Int. J. Mod. Phys. }

\newcommand{\AmS}{{\protect\the\textfont2
 A\kern-.1667em\lower.5ex\hbox{M}\kern-.125emS}}
\title{The Master Field for
the Half-Planar Approximation for Large $N$
Matrix Models and Boltzmann Field Theory}

\author{I.Ya.Aref'eva and A.P.Zubarev\thanks{Talk presented at
29 Symposium on High Energy Physics, Buckow, August, 1995} \\
  Steklov Mathematical Institute,
  Vavilov 42, GSP-1, 117966, Moscow, Russia }

\begin{document}
\pagestyle{myheadings}

\begin {abstract}
In this talk results of study  in various  dimensions
of the Boltzmann master field for a subclass of planar diagrams,
so called half-planar diagrams,
found in the recent work by Accardi, Volovich and one of us (I.A.)
are presented.
\end {abstract}

\maketitle

\section{Introduction}

The problem of analytic summation of all planar diagrams in
realistic models is still unsolved.
Its solution is closely related with problem of finding the leading
asymptotics in $N \times N$ matrix models for large $N$ and may have
important applications to the hadron dynamics  \cite {tH,Ven,Wit,AS}.
In the  early 80-s it was suggested  \cite {Wit} that there exists a master
field  which dominates in the large $N$ limit of  invariant correlation
functions of a matrix field.

The problem of construction of the master field has been
discussed in many works, see for example \cite {Haan}-\cite {GH}.
Gopakumar and Gross \cite {GG} and Douglas \cite {Doug2}
have constructed the master field for an arbitrary matrix model
in terms of correlation functions.
There has been a problem of construction an operator realization
for the master field without knowledge of correlation functions.
Recently this problem has been solved in \cite{AV}
and it was shown that the master fields satisfy to standard equations of
relativistic field theory but fields are quantized according to a new rule.

In this talk we are going to demonstrate that an operator realization
for the master field for a subset of  planar diagrams, so called half-planar
(HP) diagrams, proposed in \cite{AAV} gives an  analytical summations
of HP diagrams (see \cite{AZ} fore more details).
 This construction  deals with  the  master field
in a  modified interaction representation in the free (Boltzmannian)
Fock space. This new interaction representation involves not the ordinary
exponential function  of the interaction but a rational function of the
interaction. Corresponding correlation functions  satisfy the  Boltzmannian
Schwinger-Dyson equations which are simpler than the usual Schwinger-Dyson
equations. In particular in the case of quartic interaction one has a
closed set of equations  for two- and four-point correlation functions.
We solve explicitly  this system of equations.
In the case of $D$-dimensional space-time we get a Bethe-Salpiter-like
equation for the four-point correlation function.
A special approximation reduces this system  of integral equations
to a linear integral equation which was considered \cite{Rothe}
in the rainbow approximation in the usual field theory.

A solution of the Boltzmannian Schwinger-Dyson equations
can be considered as a first non-trivial approximation to the planar
correlation functions.  Note in this context that in all previous attempts
of approximated treatment of planar theory were used some non-perturbative
approximation \cite {Sl,IA,Fer}.  Topologically  diagrams  representing
the perturbative series of Boltzmann correlation functions  look as
half-planar diagrams of the usual diagram technique for matrix models
\cite{AAV}.  We compare numerically the two- and  four-point HP
correlation functions with the corresponding planar correlation functions
for the one matrix model.  For large variety of coupling constant the HP
approximation reproduces the planar approximation with good accuracy.
This fact gives us an optimism and makes sensitive a study of the
HP approximation for realistic models.

\section{Half-Planar Approximation for the One Mat\-rix Model}
The master field  in zero dimensional case
is defined as
$ \phi =a + a^+  $,
where $a$ and $a^+$ satisfy the following relation
\begin{eqnarray}
\label{1}
 aa^+=1.
\end{eqnarray}
This algebra has a realization in the free (or Boltzmannian) Fock space
\cite {OWG} generated by the vacuum
$|0\rangle $,
$a|0\rangle =0$,
and $n$-particle states
$|n \rangle=$
$(a^+)^n|0\rangle.$

A free $n$-point Green's function is defined as the vacuum expectation
of $n$-th power of master field
\begin{eqnarray}
\label{4}
G^{(0)}_n=\langle 0|\phi ^n |0\rangle .
\end{eqnarray}
As it is well-known, the Green's function (\ref{4}) is  given by a $n$-th
moment of Wigner's distribution  \cite{BIPZ,Voi}
\begin{eqnarray}
G^{(0)}_{2n}=\int_{-2}^{2}\frac{d \lambda }{2 \pi }
\lambda ^{2n}\sqrt{4-\lambda ^2}  =
\frac{(2n)!}{n!(n+1)!}.
\nonumber
\end{eqnarray}
This representation
can be also obtained as a solution of the Schwinger-Dyson equations
\begin{eqnarray}
G^{(0)}_{2n}=\sum_{m=1}^{n}G^{(0)}_{2m-2}G^{(0)}_{2n-2m}.
\nonumber
\end{eqnarray}

Interacting Green's functions are defined by the  formula \cite{AAV}
\begin{eqnarray}
\label{16}
G_n=\langle 0|\phi ^n
(1+S_{int}(\phi ))^{-1}
|0\rangle .
\end{eqnarray}
In contrast to the ordinary quantum field theory
where one deals with the exponential function of an interaction,
here we deal with  the
 rational function of an interaction.
In \cite{AZ}  it was shown that under natural assumptions  the form
(\ref{16})  is unique one
which admits  Schwinger-Dyson-like equations.

For the case of quartic interaction $S_{int}=g\phi ^4$
the Boltzmannian Schwinger-Dyson equations have the form
\begin{eqnarray}
G_n=
\sum _{l=1}^{k-1}
G^{(0)}_{k-l-1}G_{l+n-k-1}+
\label{23}
\end{eqnarray}
\begin{eqnarray}
\sum _{l=k+1}^{n}G^{(0)}_{l-k-1}G_{n+k-l-1}
-g[G_{n-k}G_{k+2}+
\nonumber
\end{eqnarray}
\begin{eqnarray}
G_{n-k+1}G_{k+1}+
G_{n-k+2}G_{k}+
\nonumber
\end{eqnarray}
\begin{eqnarray}
G_{n-k+3}G_{+k-1}].
\nonumber
\end{eqnarray}
The distinguish feature of  equations (\ref{23}) is that for $n \ge
4$ and $2\le k\le n-1 $ the right hand side of (\ref{23}) does not contain
the  Green's functions $G_m$ with $m > n$.  This fact permit us to write
down a closed set of equations for any $G_2$
and $G_4$
\begin{eqnarray}
G_2=1-gG_2G_2-gG_4,
\nonumber
\end{eqnarray}
\begin{eqnarray}
G_4=2G_2-2gG_2G_4.
\label{27}
\end{eqnarray}
These equations follow from (\ref{23}) for $n=2$, $k=1$ and
$n=4$, $k=2$. We set $G_0=1$ to exclude  vacuum insertions. From
(\ref{27}) one gets
\begin{eqnarray}
\label{28}
2g^2G^3_2+3gG^2_2+G_2-1=0.
\end{eqnarray}
Using the Cordano formula for $g\le \frac{\sqrt{3}}{18}$ we
have
\begin{eqnarray}
G_{2}= \frac{1}{\sqrt{3}g}
{\rm Re}e^{i\frac{\pi}{6}}(\sqrt{1-108g^2}-
i 6\sqrt{3}g)^{\frac{1}{3}}-
\nonumber
\end{eqnarray}
\begin{eqnarray}
\frac{1}{2g}=1-3g+16g^2-105g^3+768g^4-...~.~~
\nonumber
\end{eqnarray}
Green's functions  for $n>4$ are also determined from
(\ref{23}).

\begin{table*}[hbt]
\begin{tabular*}{\textwidth}{@{}l@{\extracolsep{\fill}}llllllll}
\hline
$g$ & $10^{-3}$ & $10^{-2}$ & $10^{-1}$ & $1$ & $10$ & $10^2$&$10^3$
\\
\hline
$\Pi _2$
& $9.98\cdot  10^{-1}$
& $9.81 \cdot 10^{-1}$
& $8.6 \cdot 10^{-1}$
& $5.2 \cdot 10^{-1}$
& $2.1 \cdot 10^{-1}$
& $7.4 \cdot 10^{-2}$
& $2.4 \cdot 10^{-2}$
\\
$G_2$
& $9.97\cdot  10^{-1}$
& $9.71 \cdot 10^{-1}$
& $8.0 \cdot 10^{-1}$
& $4.0 \cdot 10^{-1}$
& $1.3 \cdot 10^{-1}$
& $3.2 \cdot 10^{-2}$
& $7.4 \cdot 10^{-3}$
\\
$\Pi _4$
& $1.99$
& $1.96$
& $1.42$
& $4.84 \cdot 10^{-1}$
& $7.87 \cdot 10^{-2}$
& $9.26 \cdot 10^{-3}$
& $9.76 \cdot 10^{-4}$
\\
$G_4$
& $1.99$
& $1.91$
& $1.38$
& $4.43 \cdot 10^{-1}$
& $7.16 \cdot 10^{-2}$
& $8.65 \cdot 10^{-3}$
& $9.37 \cdot 10^{-4}$
\\
\hline
\end{tabular*}
\end{table*}

Now let us compare the Boltzmann
theory and the planar approximation for the one-matrix model.
Green's functions for the one-matrix model in the planar approximation
are defined as
\begin{eqnarray}
\Pi _{2n}(g) =\lim _{N\to \infty }\frac{1}{N^{1+n}}
\frac{1}{{\cal Z}}\int DM \tr (M^{2n})
\nonumber
\end{eqnarray}
\begin{eqnarray}
exp[-\frac{1}{2}tr (M^2)-
\frac{g}{4N}\tr (M^4)],~~~~
\label{mm}
\end{eqnarray}
where ${\cal Z}$ is a normalization factor.
The integration in (\ref {mm}) is over $N\times N$ hermitian matrices.
According the 't Hooft diagram technique the perturbative expansion
in the coupling constant of the correlation functions (\ref {mm})
is represented by a sum of all planar double-line graphs  \cite {tH}.
Due to the normalization
factor $N^{-(1+n)}$ external lines corresponding to
$\tr M^{2n}$ can be treated as lines of a generalized vertex.
 We shall call two
double-line  planar graphs
topologically equivalent  if one of them can be transformed into the
other by a continuous deformation on the plane.

A  planar non-vacuum graph is  an
HP graph if it is topologically equivalent to a graph
which can  be drawn  so that all its vertices lie on some plane
line  in the right of the
generalized vertex $\tr M^{2n}$ and all propagators lie in the upper-half
 plane without
overlapping  \cite {AAV}.
Also we shall call a planar graph  HP-irreducible if it is represented as
an HP graph in an unique way.
A simple analysis shows that  an HP graph is HP-irreducible if it does not
contain any tadpole subgraphs.
By a graph
with a tadpole we mean a graph  with a subgraph
which contains two lines coming from the same vertex
so that after removing of these two lines the remaining subgraph becomes
disconnected with the rest of the graph.

In \cite{AAV} it has been shown  that if one considers
some special approximation
for the planar theory, namely so called HP approximation, then
Green's functions of the one-matrix model in this approximation
coincide with correlation functions in a corresponding  Boltzmann
theory,
\begin {eqnarray} 
\Pi ^{appr} _{2n}(g)=<\phi ^{2n}(1+g\phi ^{4})^{-1}>.
							  \label {ahp}
\end   {eqnarray} 
In the LHS we select only  irreducible HP graphs
(this approximation we call the  HP approximation)
and in the RHS we omit all graphs with tadpoles.
Let us sketch a pure combinatorial proof of equation (\ref {ahp}) \cite{AZ}.
It is based on  the following two statements.
The first one states that  the planar correlation functions
without vacuum insertions in all order of perturbation theory are
represented by a
sum of all topologically non-equivalent graphs without
any combinatorial factors.
According the second one,  all graphs
in the Boltzmann theory contribute into
correlations functions (\ref {16}) without any combinatorial factors.

Now let us compare numerically the HP approximation
with the planar approximation.
The explicit formulas for an arbitrary planar
Green's functions  are well known  \cite{BIPZ}.
On the Table  we give
 the results of numerical calculations of the HP Green's functions
$G_2$, $G_4$ and
the planar Green's function $\Pi _2$, $\Pi _4$
for the same values of the coupling constant $g$.
One can see that the answers for HP Green's functions
$G_2$, $G_4$ practically saturate the planar Green's functions
$\Pi _2$, $\Pi _4$ in the board interval of the values of $g$.

It is instructive to reformulate the Boltzmann theory so
that it  reproduces only tadpole-free HP graphs.
For this purpose let us defined the fields $\psi$ and $\phi$ as follows
$\phi = a+a^+,$
$\psi = b + b^+,$
where $a$, $a^+$, $b$ and $b^+$ satisfy the following relations
\begin{eqnarray}
a a^+=1,~~ b b^+=1,~~a b^+=0,~~b a^+=0.
\nonumber
\end{eqnarray}
This algebra has  realization in the Boltzmannian Fock space under
the vacuum
$|0\rangle : ~$
$a|0\rangle= $
$b|0\rangle=0 $.

\begin{figure*}
\begin{center}
\unitlength=1.00mm
\special{em:linewidth 0.4pt}
\linethickness{0.4pt}
\begin{picture}(142.00,34.00)
\put(5.00,30.00){\makebox(0,0)[cc]{Figure 1}}
\put(123.00,10.84){\oval(20.00,14.33)[t]}
\put(133.00,10.67){\circle*{4.67}}
\put(31.67,28.67){\circle*{4.67}}
\put(123.00,28.84){\oval(20.00,10.33)[t]}
\emline{133.00}{28.67}{1}{138.33}{34.00}{2}
\emline{113.00}{28.67}{3}{107.67}{34.00}{4}
\put(133.00,28.67){\circle*{4.67}}
\emline{70.00}{33.67}{5}{75.00}{28.67}{6}
\emline{75.00}{28.67}{7}{80.00}{33.67}{8}
\emline{26.67}{33.67}{9}{31.67}{28.67}{10}
\emline{31.67}{28.67}{11}{36.67}{33.67}{12}
\put(53.33,10.67){\makebox(0,0)[cc]{$=$}}
\put(95.00,10.67){\makebox(0,0)[cc]{$+$}}
\put(95.00,28.67){\makebox(0,0)[cc]{$+$}}
\put(53.00,28.67){\makebox(0,0)[cc]{$=$}}
\put(24.00,10.67){\circle*{1.00}}
\put(42.00,10.67){\circle*{1.00}}
\put(66.00,10.67){\circle*{1.00}}
\put(84.00,10.67){\circle*{1.00}}
\put(105.00,10.67){\circle*{1.00}}
\put(141.00,10.67){\circle*{1.00}}
\put(140.33,7.67){\makebox(0,0)[cc]{$_y$}}
\put(105.00,7.67){\makebox(0,0)[cc]{$_x$}}
\put(84.00,7.67){\makebox(0,0)[cc]{$_y$}}
\put(66.00,7.67){\makebox(0,0)[cc]{$_x$}}
\put(42.00,7.67){\makebox(0,0)[cc]{$_y$}}
\put(24.00,7.67){\makebox(0,0)[cc]{$_x$}}
\put(66.00,10.67){\line(1,0){18.00}}
\put(105.00,10.67){\line(1,0){8.00}}
\put(113.00,10.67){\line(0,0){0.00}}
\put(122.50,10.50){\oval(19.00,7.00)[t]}
\linethickness{1.5pt}
\put(142.00,28.67){\line(-1,0){37.00}}
\put(84.00,28.67){\line(-1,0){18.00}}
\put(25.00,28.67){\line(1,0){13.00}}
\put(113.00,10.67){\line(1,0){20.00}}
\put(24.00,10.67){\line(1,0){18.00}}
\put(133.00,10.67){\line(1,0){8.00}}
\end{picture}
\end{center}
\label{fig1}
\vspace{-10mm}
\end{figure*}

Now let us consider the following Green's functions
\begin{eqnarray}
F_{n} = \langle 0 | \psi  \phi ^{n-2} \psi  (1+S_{int})^{-1}|0\rangle,
\nonumber
\end{eqnarray}
\begin{eqnarray}
S_{int}=
g\psi :\phi \phi : \psi=
g\psi \phi \phi  \psi -
g\psi \psi.
\nonumber
\end{eqnarray}
The Schwinger-Dyson equations for
the Green's functions $F_2$ and $F_4$
have the form
\begin{eqnarray}
F_2=1-gF_4+g F_2,~
F_4=-g F_2 F_4 + F_2.
\nonumber
\end{eqnarray}

 From these equations  we  find
\begin{eqnarray}
F_2=\frac{-1+g +\sqrt{1+2g-3g^2}}{2g(1-g)},
\nonumber
\end{eqnarray}
\begin{eqnarray}
F_4=\frac{1+g -\sqrt{1+2g-3g^2}}{2g^2}.
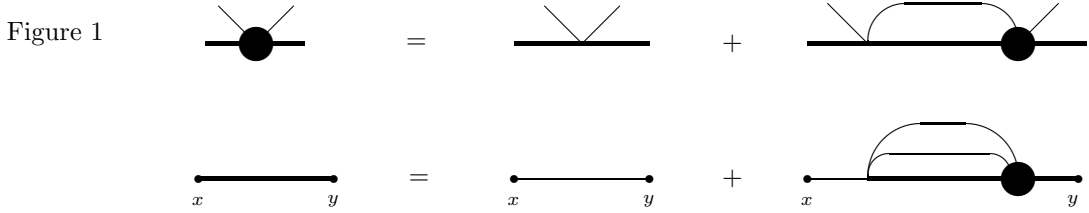
\nonumber
\end{eqnarray}

\section{Boltzmann Correlation Functions for $D$-Dimensional Space-Time}

In this section we  derive the Schwinger-Dyson equations for Boltzmann
correlation functions in $D$-dimensional Euclidean space.
To avoid  problems with tadpoles let us consider the two-field formulation.
We adopt the following notations. Let
$\psi (x)=\psi ^+(x)+\psi ^-(x),$
$\phi (x)=$$\phi ^+(x)+$$\phi ^-(x)$
be the Bolzmann fields with creation and annihilation operators
satisfying the relations
\begin{eqnarray}
\psi ^-(x) \psi ^+(y)=
\phi ^-(x) \phi ^+(y)=D(x,y),
\nonumber
\end{eqnarray}
\begin{eqnarray}
\psi ^-(x) \phi ^+(y)=\phi ^-(x) \psi ^+(y)=0,
\nonumber
\end{eqnarray}
where
$
D(x,y)=\int \frac{d^Dk}{(2\pi ) ^D}(k^2+m^2)^{-1}e^{ik(x-y)}  $
is $D$-di\-men\-sio\-nal Eucli\-dean  propagator.
The $n$-point Green's function is defined by
\begin {eqnarray} 
 F_{n}(x_1,...,x_n)=
\langle 0|\psi (x_1)\phi (x_2)...
\label {5.0}
\end{eqnarray}
\begin{eqnarray}
\phi (x_{n-1})\psi (x_n)
 (1+\int d^{D}x g \psi :\phi \phi : \psi))^{-1}
|0\rangle .
\nonumber
\end   {eqnarray} 

Let us write down  the Schwinger-Dyson equations
for the two- and four-point correlation functions.
We have
 \begin{eqnarray}
(-\bigtriangleup +m^{2})_{x}F_{2}(x,y)=gD(x,x)F_{2}(x,y)
\nonumber
\end{eqnarray}
\begin{eqnarray}
-gF_{4}(x,x,x,y)+\delta (x-y) ,
						\label {2p}
\end{eqnarray} 
\begin{eqnarray} 
(-\bigtriangleup +m^{2})_{y}F_{4}(x,y,z,t)=
\nonumber
\end{eqnarray}
\begin{eqnarray}
 -gF_{4}(y,y,z,t)F_2(x,y)+\delta (y-z)F_2(x,t).
						 \label{4p}
\end{eqnarray} 
Here we also assume that all vacuum insertions are dropped out.
We see that equation
(\ref{4p})  does not contain six-point correlation functions.
As a consequence we have a closed set of  equations which are
enough to find $F_{2}$ and $F_{4}$.
We define an one-particle irreducible  (1PI) 4-point function
$\Gamma_{4}(x,y,z,t)$  as
\begin {eqnarray} 
\Gamma_{4}(x,y,z,t)=
\int
dx^{\prime}dy^{\prime}dz^{\prime}
dt^{\prime}
F^{-1}_{2}(x,x^{\prime})
\label {ppp}
\end{eqnarray}
$D^{-1}(y,y^{\prime})
D^{-1}(z,z^{\prime})
F^{-1}_{2}(t,t^{\prime}){\cal F}_{4}(x^{\prime},
y^{\prime},z^{\prime},t^{\prime}),~~~$
where
${\cal F}_{4} $ is a connected
part of $F_{4}$
$$F_{4}(x,y,z,t)={\cal F}_{4}(x,y,z,t)+
F_{2}(x,t)D(y,z).~~~~$$

Note that in the contrast to the usual case in the RHS of
(\ref {ppp}) we multiply ${\cal F}_{4}$ only on two full 2-point
Green functions while in the usual case to get an 1PI Green function one
multiplies  an $n$-point
Green function on $n$ full 2-point functions. From (\ref {2p}) and
(\ref {ppp}) we have
\begin {eqnarray} 
\Gamma_{4}(p,k,r)= -g-
g\int dk^{\prime}
F_{2}(p+k-k^{\prime}) \times
\nonumber
\end{eqnarray}
\begin{eqnarray}
D(k^{\prime})\Gamma_{4}(p+k-k^{\prime},
k^{\prime},r)
							  \label {bsl}
\end   {eqnarray} 

Equation (\ref {bsl}) is the Bethe-Salpeter-like equ\-ation
with the kernel which contains an unknown  function $F_{2}$.
As in the usual case we can write down $F_{2}$ in term of
the self-energy
function $\Sigma _{2}$
\begin {eqnarray} 
F_{2}=\frac{1}{p^{2}+m^{2}+\Sigma _{2}}
\nonumber
\end   {eqnarray} 
and write equation (\ref {2p}) as an equation for $\Sigma _{2}$,
\begin {eqnarray} 
\Sigma _{2} (p)=g\int dkdq F_{2}(k)D(q)
\times
\nonumber
\end{eqnarray}
\begin{eqnarray}
\label {si4}
D(p-k-q)\Gamma _{4}(p,k,q).
\end   {eqnarray} 

Equation (\ref {si4}) is similar to the  usual
relation between the self-energy function $\Sigma _{2}$ and the
4-point vertex function for $\varphi ^{4}$ field theory,
meanwhile equation (\ref{bsl}) is specific for the
Boltzmann field theory.
Equations (\ref{bsl}) and (\ref{si4}) are drawn on Fig.1.

\begin{figure*}
\begin{center}
\unitlength=1.00mm
\special{em:linewidth 0.4pt}
\linethickness{0.4pt}
\begin{picture}(145.00,51.00)
\put(5.00,45.00){\makebox(0,0)[cc]{Figure 2}}
\put(37.50,25.17){\oval(35.00,32.33)[t]}
\emline{20.00}{25.00}{1}{12.33}{32.67}{2}
\emline{55.00}{25.00}{3}{63.00}{33.00}{4}
\put(25.00,20.00){\dashbox{0.67}(25.00,17.00)[cc]{}}
\bezier{200}(103.00,25.00)(120.00,43.00)(137.00,25.00)
\bezier{248}(103.00,25.00)(120.00,51.00)(137.00,25.00)
\bezier{164}(30.00,25.00)(37.67,44.00)(45.00,25.00)
\bezier{100}(30.00,25.00)(37.67,35.00)(45.00,25.00)
\bezier{76}(117.00,25.00)(120.00,34.00)(123.00,25.00)
\bezier{40}(117.00,25.00)(120.00,29.00)(123.00,25.00)
\put(115.00,25.00){\dashbox{0.67}(10.00,6.33)[cc]{}}
\put(125.00,22.00){\dashbox{0.67}(16.00,3.00)[cc]{}}
\put(101.00,22.00){\dashbox{0.67}(14.00,3.00)[cc]{}}
\put(101.00,25.00){\dashbox{0.67}(40.00,16.33)[cc]{}}
\put(120.00,15.00){\makebox(0,0)[cc]{$b)$}}
\put(37.67,15.00){\makebox(0,0)[cc]{$a)$}}
\linethickness{0.8pt}
\put(10.00,25.00){\line(1,0){55.00}}
\put(95.00,25.00){\line(1,0){50.00}}
\end{picture}
\end{center}
\label{fig2}
\vspace{-20mm}
\end{figure*}

Equations for $\Gamma _4$ and $\Sigma _2$ contain diver\-gen\-ces.
To re\-mo\-ve them we ap\-ply
to Boltz\-man\-nian Green's functions (\ref{5.0})
the standard renormalization procedure based on
$R$-operation.  The perturbative expansion of the Boltzmannian
Green's functions (\ref {5.0})  is represented by the sum of HP
Feynman graphs.
We draw $\phi $-propagator by thin lines and $\psi $-propagator by
thick lines. Only two- and four-point graphs may be divergent.
Any two-point subgraph
of the HP graph is also from the set of HP
graphs and performing contractions of the two-point subgraphs
of the HP graph one gets again an HP graph.
But four-point subgraphs of an HP graph may be of HP type
and may be not.
We refer to the later case as to the case of $\Pi $-subgraphs.
Examples of divergent HP-subgraphs and $\Pi$-subgraphs are drawn
on Fig. $2a$ and $2b$, respectively.

A detailed consideration \cite{AZ} shows that all divergent parts can be
collected  to counterterms. Four-point
divergent parts of $\Pi $-type require a counterterm
$ :\psi \psi :  \psi ^-\psi ^- $.

Let us assume dimensional regularization and the minimal subtraction scheme
with a scale $\mu$. For renormalized correlation functions  we have
\begin {eqnarray} 
F_n(x_1,...,x_n;  M(\mu),g(\mu),\lambda (\mu),\mu)=
\nonumber
\end{eqnarray}
\begin{eqnarray}
\langle 0|\psi (x_1)\phi (x_2)...\phi (x_{n-1})\psi (x_n)
\times
\nonumber
\end{eqnarray}
\begin{eqnarray}
(1+\int d^{D}x[{\cal L}_{int}(\psi , \phi )
+{\cal L}_{ct}(\psi, \phi )])^{-1}
|0\rangle ,
\nonumber
\end   {eqnarray} 
where
\begin{eqnarray}
{\cal L}_{int}= (M^2-m^{2})\psi ^2 +
\nonumber
\end{eqnarray}
\begin{eqnarray}
\mu ^{\varepsilon}g\psi :\phi \phi : \psi +
\mu ^{\varepsilon} \lambda :\psi \psi :  \psi ^- \psi ^-,
\nonumber
\end{eqnarray}
\begin{eqnarray}
{\cal L}_{ct}=
(Z_{\psi}-1)
(\partial  _{\mu} \psi )^{2}+
\delta M^2\psi ^2+
\nonumber
\end{eqnarray}
\begin{eqnarray}
\mu ^{\varepsilon}\delta g\psi :\phi \phi :  \psi+
\mu ^{\varepsilon}\delta \lambda :\psi \psi:  \psi ^- \psi ^-
\nonumber
\end{eqnarray}
and $\varepsilon =4-D$. Here we add a new term to the interaction.
This $\lambda$-term modifies only the mass term in equation (\ref {2p})
(see  \cite {AZ} for more details).

The renormalized correlations functions
satisfy  the following renormalization group equation
\begin{eqnarray}
(\mu \frac{\partial}{\partial \mu}+
\beta _{g}\frac{\partial}{\partial g}+
\beta _{\lambda}\frac{\partial}{\partial \lambda}-
\gamma _M  \frac{\partial}{\partial \ln M^2}+
 \gamma )\times
\nonumber
\end{eqnarray}
\begin{eqnarray}
F_n(x_1,...,x_n;M(\mu),g(\mu),\lambda (\mu),\mu)=
0,
					       \label{5.23}
\end{eqnarray}
where
\begin{eqnarray}
\beta _{g}=  \mu \frac{ \partial g(\mu )}{ \partial \mu} ,~~
\beta _{\lambda}=  \mu \frac{\partial \lambda (\mu )}{ \partial \mu} ,
\nonumber
\end{eqnarray}
\begin{eqnarray}
\gamma _M =  -\mu M^{-2} \frac{\partial M^2(\mu )}{ \partial \mu} ,~~
\gamma =  \mu \frac{\partial \ln Z_{\psi }(\mu )}{ \partial \mu } .
\nonumber
\end{eqnarray}
The difference between  (\ref{5.23}) and the usual re\-nor\-malization
group equation is that the ano\-malous dimension $\gamma $ in (\ref {5.23})
is not multiplied on $n/2$.  There are also a differences in the
expressions for the beta-functions in terms of the counterterms $\delta
g$,  $\delta \lambda$ and $Z_{\psi } - 1 $. In dimensional regularization
the counterterms are poles in $\varepsilon $,
\begin{eqnarray} \delta
g=\sum_{i=1}^{\infty}\frac{a_i(g, \lambda)}{\varepsilon ^i},~~
\delta\lambda=\sum_{i=1}^{\infty}\frac{b_i(g,\lambda)}{\varepsilon ^i},
\nonumber
\end{eqnarray}
\begin{eqnarray}
Z_{\psi}=1+\sum_{i=1}^{\infty}\frac{c_i(g, \lambda)}{\varepsilon ^i}.
\label{5.24}
\end   {eqnarray}

As in the usual case one can write the low-order expression
for the $\beta$-function in terms of $c_{1}$ and $a_{1}$.
But now there is a new expression for $g_{0}$ in terms of
$\delta g$  and $Z_{\psi} $. We have
\begin {eqnarray} 
g_0=\mu ^{\varepsilon}(g+\delta g)Z_{\psi} ^{-1},
\nonumber
\end{eqnarray}
\begin{eqnarray}
\lambda_0=\mu ^{\varepsilon}(\lambda+\delta \lambda)Z_{\psi} ^{-2}.
							  \label {5.25}
\end   {eqnarray} 
Applying $\mu \frac{d}{d\mu}$ to both sides of (\ref {5.25}) one
finds
\begin{eqnarray}
\beta _g(g)=(1-\frac{\partial}{\partial \ln g})(g c_1(g, \lambda)-
a_1(g, \lambda)).
\nonumber
\end{eqnarray}
\begin{eqnarray}
\beta _{\lambda}(\lambda)=(1-\frac{\partial}{\partial \ln \lambda})
(2\lambda c_1(g,\lambda)-b_1(g,\lambda)).
\nonumber
\end{eqnarray}
Taking into account the explicit form of $c_1(g,\lambda)$, $a_1(g, \lambda)$
and $b_1(g,\lambda )$
 we get
\begin {eqnarray} 
\beta _g(g, \lambda)=
\frac{g^2}{8 \pi ^2}+
 \frac{g^3}{(16\pi ^2)^2}+O( g^4,~\lambda ^4),
\nonumber
\end{eqnarray}
\begin{eqnarray}
\beta _{\lambda}(g, \lambda)=
\frac{\lambda ^2}{8\pi ^{2}}+
O( g^4,~\lambda ^4).
\end   {eqnarray} 

Note that there is  only a  numerical difference of  the HP
$\beta$ function with the standard $\beta$  function.

$$~$$
{\bf ACKNOWLEDGMENT}

\vspace{5mm}

I.A. thanks D.Lust, D.Ebert and In\-sti\-tu\-te of Phy\-sics of
Hum\-boldt University in Berlin whe\-re a part of this work
was done for the kind hospitality. I.A. is supported in
part by Deut\-sche Forschungsgemeinschaft under project
DFG 436 RUS 113/29. Both authors are supported by RFFR
grant 93011147 and ISF grant M1L300.
We are grateful to P.Medvedev and I.Volovich for useful
discussions.
$$~$$


\end{document}